# Growth and characterization of quaternary-alloy ferromagnetic semiconductor (In,Ga,Fe)Sb

Tomoki Hotta,[1,a)] Kengo Takase,[1] Kosuke Takiguchi[1], Karumuri Sriharsha[1], Le Duc Anh,[1,2,3,b)] and Masaaki Tanaka[1,4,c)]

[1]*Departmen of Electrical Engineering & Information Systems, The University of Tokyo, 7-3-1 Hongo, Bunkyo, Tokyo 113-8656, Japan.*
[2]*Institute of Engineering Innovation, The University of Tokyo, 7-3-1 Hongo, Bunkyo, Tokyo 113-8656, Japan.*
[3]*PRESTO, Japan Science and Technology Agency (JST), 4-1-8 Honcho Kawaguchi, Saitama 332-0012, Japan.*
[4]*Center for Spintronics Research Network (CSRN), The University of Tokyo, 7-3-1 Hongo, Bunkyo, Tokyo 113-8656, Japan.*

Email: a) hotta@cryst.t.u-tokyo.ac.jp
b) anh@cryst.t.u-tokyo.ac.jp
c) masaaki@ee.t.u-tokyo.ac.jp

**Abstract**

We study the growth and properties of quaternary-alloy ferromagnetic semiconductor (FMS) $(In_{0.94-x},Ga_x,Fe_{0.06})Sb$ ($x = 5\% - 30\%$, Fe concentration is fixed at 6%) grown by low temperature molecular beam epitaxy (LT-MBE). Reflection high-energy electron diffraction (RHEED) patterns, scanning transmission electron microscopy (STEM) lattice images, and X-ray diffraction (XRD) spectra indicate that the $(In_{0.94-x},Ga_x,Fe_{0.06})Sb$ layers have a zinc-blende crystal structure without any other second phase. The lattice constant of the $(In_{0.94-x},Ga_x,Fe_{0.06})Sb$ films changes linearly with the Ga concentration $x$, indicating that Ga atoms substitute In atoms in the zinc-blend structure. We found that the carrier type of $(In_{0.94-x},Ga_x,Fe_{0.06})Sb$ can be systematically controlled by varying $x$, being n-type when $x \leq 10\%$ and p-type when $x \geq 20\%$. Characterizations using magnetic circular dichroism (MCD) spectroscopy indicate that the $(In_{0.94-x},Ga_x,Fe_{0.06})Sb$ layers have intrinsic ferromagnetism with relatively high Curie temperatures ($T_C = 40 - 120$ K). The ability to widely control the fundamental material properties (lattice constant, bandgap, carrier type, magnetic property) of $(In_{0.94-x},Ga_x,Fe_{0.06})Sb$ demonstrated in this work is essential for spintronic device applications.

**Main**

Ferromagnetic semiconductors (FMSs), which are semiconductors doped with a large amount of magnetic elements, are promising materials for device applications because they possess both semiconducting and ferromagnetic properties. In the past three decades, Mn-doped III-V FMSs, such as (In,Mn)As[1], (Ga,Mn)As[2–4] and (In,Ga,Mn)As[5–7], have been extensively studied as prototypical FMSs. However, these Mn-doped FMSs show Curie temperature ($T_C$) much lower than room temperature and are only p-type. These shortcomings severely hinder their potential applications in practical spintronics devices, especially those based on ferromagnetic p-n junctions. One promising alternative approach, recently developed by our group, is the Fe-doped III-V FMSs, such as n-type (In,Fe)As[8–12], p-type (Ga,Fe)Sb[13–16], p-type (Al,Fe)Sb[17] and n-type (In,Fe)Sb[18–20]. These new FMSs exhibit strong ferromagnetism, some with $T_C$ much higher than 300 K, and both n-type and p-type FMSs can be realized. These important features overcome most of the inherent problems of the Mn-doped III-V FMSs and thus make the Fe-doped FMSs promising candidates for spintronic materials.

Among the Fe-doped III-V FMSs, (Ga,Fe)Sb and (In,Fe)Sb are the two most interesting materials: While (Ga,Fe)Sb is p-type and has perpendicular magnetic anisotropy[21–23], (In,Fe)Sb is n-type and has in-plane magnetic anisotropy[20]. Both of them, however, exhibit room-temperature ferromagnetism when being doped with high enough Fe concentration (> 15 %). These features thus ignite great interest in the quaternary-alloy (In,Ga,Fe)Sb, which fills the gap between the n-type FMS (In,Fe)Sb and p-type FMS (Ga,Fe)Sb. In non-magnetic (In$_{1-x}$,Ga$_x$)Sb thin films, by changing the Ga concentration $x$, the lattice constant and band gap were reported to change from 0.648 nm and 0.18 eV ($x = 0$, InSb) to 0.61 nm and 0.78 eV ($x = 1$, GaSb), respectively[24]. Therefore, we expect that the quaternary-alloy (In,Ga,Fe)Sb potentially can be a versatile ferromagnetic semiconductor, in which various physical properties such as the lattice constant, band gap, carrier type, and magnetic anisotropy, can be controlled by changing the Ga content. In particular, the evolution of $T_C$ and magnetic anisotropy with the Ga content is of interest because they are determined by the band structure and epitaxial strain of the FMS thin film. Such a highly tunable FMS platform is essential for realizing practical spintronic devices.

In this work, we conducted a systematic study on the epitaxial growth, crystal structure, transport and magnetic properties of (In$_{0.94-x}$,Ga$_x$,Fe$_{0.06}$)Sb thin films, where we fixed the Fe concentration at 6% and varied the Ga concentration $x$ ($x = 5\%, 10\%, 20\%,$ and $30\%$). This series of samples were grown on semi-insulating GaAs (001) substrates by low-temperature molecular beam epitaxy (LT-MBE). The schematic sample structure and material parameters are shown in Fig. 1(a) and Table I, respectively. After evaporating the native oxide surface layer of the GaAs substrate at 580°C, a 100-nm-thick GaAs buffer layer was grown at 540°C followed by a 6-nm-thick AlAs buffer layer at the same temperature. Then, we grew a 100-nm-thick AlSb buffer layer at 470°C to relax the large lattice mismatch (~ 7 – 13%) between the substrate and the $(In_{0.94-x},Ga_x,Fe_{0.06})Sb$ layer and to obtain a smooth growth interface. After cooling the substrate temperature to 250°C, an $(In_{0.94-x},Ga_x,Fe_{0.06})Sb$ layer with a thickness of from 10 to 15 nm was grown at a rate of 0.5 μm/h. Finally, a 2-nm-thick (In,Ga)Sb cap layer was grown to prevent oxidation of the $(In_{0.94-x},Ga_x,Fe_{0.06})Sb$ layer. As shown in Fig. 1(b), *in-situ* reflection high energy electron diffraction (RHEED) patterns taken along the [$\bar{1}10$] azimuth during the MBE growth of the $(In_{0.94-x},Ga_x,Fe_{0.06})Sb$ layers are bright and streaky, indicating good layer-by-layer growth of thin films with a zinc-blende crystal structure. Then we characterized the crystal structure of the $(In_{0.94-x},Ga_x,Fe_{0.06})Sb$ sample with $x = 30\%$ using a scanning transmission electron microscopy (STEM) lattice image and a transmission electron diffraction (TED) pattern, as shown in Fig. 1(c). The STEM and TED images confirm a good zinc-blende structure maintained throughout the $(In_{0.94-x},Ga_x,Fe_{0.06})Sb$ ($x = 30\%$) layer without any visible second phase, and an atomically flat interface between the top magnetic layer and the AlSb buffer layer.

Next, we characterized the crystal structure of the $(In_{0.94-x},Ga_x,Fe_{0.06})Sb$ samples using X-ray diffraction (XRD), as shown in Fig. 2. The XRD rocking curves of all samples are shown in Fig. 2(a), in which a result of an $(In_{0.94},Fe_{0.06})Sb$ ($x = 0\%$) sample is also shown as a reference. In all the XRD rocking curves, there are only peaks corresponding to the $(In_{0.94-x},Ga_x,Fe_{0.06})Sb$ layer, the AlSb buffer layer and the GaAs substrate of each sample, indicating that the $(In_{0.94-x},Ga_x,Fe_{0.06})Sb$ layers were grown without any other second phase. The peak position corresponding to the $(In_{0.94-x},Ga_x,Fe_{0.06})Sb$ (004) diffraction in each sample, estimated by fitting a Gaussian curve to the XRD data, shifts to higher angles with increasing the Ga content (inset of Fig. 2(a)). This result indicates that the out-of-plane lattice constant of the $(In_{0.94-x},Ga_x,Fe_{0.06})Sb$ layers decreases with increasing $x$. This is reasonable, because increasing $x$ is expected to decrease the intrinsic lattice constant of the bulk $(In_{0.94-x},Ga_x,Fe_{0.06})Sb$[24]. To deduce more detailed information on the lattice constant of $(In_{0.94-x},Ga_x,Fe_{0.06})Sb$ thin films, we conducted reciprocal space mapping (RSM) around the (115) reciprocal lattice point, results of which are shown in Fig. 2(b) (The peak of the $(In_{0.94-x},Ga_x,Fe_{0.06})Sb$ thin film with $x = 30\%$ cannot be detected in the RSM because of the low peak intensity). From the peak position ($Q_{[110]}$, $Q_{[001]}$) in the RSM results, pointed by cross-marks in Fig. 2(b), we estimate the in-plane lattice constant $a_\parallel$, out-of-plane lattice constant $a_\perp$, and intrinsic lattice constant $a$ of the $(In_{0.94-x},Ga_x,Fe_{0.06})Sb$ layer as $a_\parallel = \frac{\sqrt{2}}{Q_{[110]}}$, $a_\perp = \frac{5}{Q_{[001]}}$, and $a = \sqrt[3]{a_\parallel \times a_\parallel \times a_\perp}$. The intrinsic lattice constant $a$ linearly decrease with increasing $x$ obeying the Vegard's law, as can be seen in Fig. 2 (c). These results indicate that we have successfully controlled the lattice constant of the $(In_{0.94-x},Ga_x,Fe_{0.06})Sb$ layers by changing the Ga content $x$. We note that the in-plane reciprocal lattice points of the $(In_{0.94-x},Ga_x,Fe_{0.06})Sb$ layers deviate largely from that of the AlSb buffers, indicating that the lattice is partially relaxed in the top active layers. The degree of relaxation decreases with increasing $x$, because the lattice mismatch between the $(In_{0.94-x},Ga_x,Fe_{0.06})Sb$ layer and the AlSb buffer decreases from 5.1% to 3.4% when $x$ is increased from 0 to 30%. As a result, the out-of-plane epitaxial strain applied on the $(In_{0.94-x},Ga_x,Fe_{0.06})Sb$ layer, given as $\epsilon_{[001]} = \frac{a_\perp - a}{a} \times 100$ (%), increases from 1.4 to 2.2% when $x$ is increased from 0 to 20%, as shown in Fig. 2(d). In practical applications, these results are essential for accurately designing the strain effect on (In,Ga,Fe)Sb, which may affect many important properties such as magnetic anisotropy, similar to the case of $(Ga,Fe)Sb$[21], and band structure.

Next, we investigated the electrical transport properties of the $(In_{0.94-x},Ga_x,Fe_{0.06})Sb$ layers. The samples were patterned into Hall bars with a size of 200 $\mu$m × 50 $\mu$m using standard photolithography and Ar-ion milling. In the samples with $x = 5\%$, 20%, and 30%, the Hall resistances show non-linear hysteresis curves at low temperatures, which suggest the presence of ferromagnetism (Fig. 3(a)). For the sample with $x = 10$ %, the Hall measurements at temperatures lower than 300 K were difficult due to its high resistance. The carrier type and carrier concentration of the $(In_{0.94-x},Ga_x,Fe_{0.06})Sb$ layers, which were estimated from the Hall coefficients at 300 K, are summarized in Fig. 3(b) and (c). The carrier concentrations of $(In_{0.94},Fe_{0.06})Sb$ ($x = 0\%$) and $(Ga_{0.94},Fe_{0.06})Sb$ ($x = 94\%$) samples are also plotted as references. We observed a clear correlation between the Ga content $x$ and the carrier type; $(In_{0.94-x},Ga_x,Fe_{0.06})Sb$ is n-type at $x \leq 10\%$ and p-type at $x \geq 20\%$. This correlation is similar to that in non-magnetic (In,Ga)Sb samples[25,26], suggesting that the carrier characteristics are determined by the native defects of (In,Ga)Sb and rather independent of Fe. The ability to obtain both n-type and p-type by a fine tuning of the material composition is promising and important for making all-FMS PN junctions and spin devices.

We characterized the magnetic properties of our samples using magnetic circular dichroism (MCD) spectroscopy. Fig. 4(a) shows the MCD spectra of all the $(In_{0.94-x},Ga_x,Fe_{0.06})Sb$ ($x$ = 5 – 30%) samples, measured at 5 K and under a magnetic field of 1 T applied perpendicularly to the film. All the spectra, regardless of $x$, show a negative peak at 2.0 eV and a positive peak at 2.5 eV, which correspond to the critical point energies $E_1$ and $E_1+\Delta_1$, respectively, of the non-magnetic (In,Ga)Sb[27]. Moreover, there is no broad background spectrum, which would be observed if there were ferromagnetic Fe-related intermetallic precipitates in the $(In_{0.94-x},Ga_x,Fe_{0.06})Sb$ layers. These results indicate that the band structure of our $(In_{1-x-y},Ga_y,Fe_x)Sb$ layers remains the zinc-blende type of the host (In,Ga)Sb without any second phase. The MCD spectroscopy results are thus consistent with the aforementioned microscopic crystal structure characterizations. Moreover, magnetic field dependence of the MCD intensity (MCD – $H$ curves) measured at a photon energy of 2.0 eV (Fig. 4 (b)) show an open hysteresis at 5 K in all the samples, indicating the presence of ferromagnetism. We estimated $T_C$ in all the samples using Arrott plots of the MCD – $H$ curves at various temperatures. The $T_C$ values are relatively high in $(In_{0.94-x},Ga_x,Fe_{0.06})Sb$ samples with $x$ = 5% ($T_C$ = 120 K ) and $x$ = 30% ($T_C$ = 110 K), but drop to a minimum between $x$ = 10 - 20%, where a transition between p and n type conduction occurs and the samples become highly resistive. Interestingly, these $T_C$ are much higher than $T_C$ = 60 K of (In,Fe)Sb and $T_C$ = 15 K of (Ga,Fe)Sb samples with the same Fe concentration (6%).

There are some possible reasons why we obtained relatively high $T_C$ in the $(In_{0.94-x},Ga_x,Fe_{0.06})Sb$ samples despite the low Fe and carrier concentrations ($10^{17} – 10^{18}$ cm$^{-3}$). The first possible scenario is that the optimized growth conditions used in the MBE growth of the $(In_{0.94-x},Ga_x,Fe_{0.06})Sb$ samples in this study. We have set the beam-equivalent-pressure (BEP) of the Sb flux at $7.0 \times 10^{-5}$ Pa during the growth of the FMS layer, which is slightly lower than the value of $7.8 \times 10^{-5}$ Pa used in our previous works[16]. Improvement of the crystal quality is indicated by the increase of mobility (~ 70 cm$^2$/Vs) in the $(In_{0.94-x},Ga_x,Fe_{0.06})Sb$ samples, as shown in Table 1, which were usually less than 10 cm$^2$/Vs in the previous reports[16]. We also note that the carrier mobility shows a positive correlation with $T_C$ in all the $(In_{0.94-x},Ga_x,Fe_{0.06})Sb$ samples (see Table 1), suggesting an important role of itinerant carriers in mediating the ferromagnetic coupling between the Fe spins which is well known for many other FMSs[28,29] including the Fe-based III-V FMSs[11,14,19]. However, a correlation between $T_C$ and the carrier concentration in Fig. 3 (b) is not exactly positive. This suggests that there is a contribution of carrier-independent ferromagnetism in $(In_{0.94-x},Ga_x,Fe_{0.06})Sb$, which is also observed in (In,Fe)Sb[19]. Another possible reason for the enhancement of $T_C$ is local fluctuation of the Fe concentration induced by spinodal decomposition. This phenomenon was predicted theoretically[30,31] and observed in heavily Fe doped (Ga,Fe)Sb,[21] where nano-columnar-like Fe rich regions are formed throughout the sample. Comparing with the cases of (In,Fe)Sb and (Ga,Fe)Sb, the coexistence of In and Ga atoms might enhance the local fluctuation of Fe concentration in the quaternary-alloy (In,Ga,Fe)Sb. In the Fe-rich regions, where the Fe – Fe distances are short, the Fe – Fe ferromagnetic interactions are enhanced by short-range superexchange interaction, leading to a higher local $T_C$ as predicted by first-principles calculations[32]. Further elaborated studies on the local crystal structure of (In,Ga,Fe)Sb are certainly required to confirm these scenarios.

In summary, we have successfully grown a new quaternary-alloy FMS $(In_{0.94-x},Ga_x,Fe_{0.06})Sb$ ($x$ = 5 – 30%). The RHEED, STEM, and XRD characterizations indicate that the $(In_{0.94-x},Ga_x,Fe_{0.06})Sb$ layers were epitaxially grown with a zinc-blende crystal structure. The Hall measurements revealed that the carrier type of the $(In_{0.94-x},Ga_x,Fe_{0.06})Sb$ layers can be changed from n-type when $x \leq 10\%$ to p-type when $x \geq 20\%$. The MCD spectroscopy results indicate the presence of ferromagnetism in all the $(In_{0.94-x},Ga_x,Fe_{0.06})Sb$ samples with high $T_C$, which also depends on $x$. Our results indicate that the new quaternary-alloy $(In_{0.94-x},Ga_x,Fe_{0.06})Sb$ is a versatile and highly tunable FMS that is promising for semiconductor spintronics devices.

**Acknowledgments**

This work was partly supported by Grants-in-Aid for Scientific Research (18H05345, 19H05602, 19K21961, 20H05650), CREST program (JPMJCR1777) and PRESTO Program (JPMJPR19LB) of Japan Science and Technology Agency, and Spintronics Research Network of Japan (Spin-RNJ).

**Data availability**

The data that support the findings of this study are available from the corresponding author upon reasonable request.

**Table 1.** Properties of $(In_{0.94-x},Ga_x,Fe_{0.06})Sb$ samples studied in this work. We show the Ga concentration $x$, thickness $d$, carrier type, carrier concentration ($n$ or $p$), electrical resistivity $\rho$, carrier mobility $\mu$ at 300 K, and $T_C$ of all the samples examined in this work.

| $x$ (%) | $d$ (nm) | Carrier type | $n$ or $p$ (cm$^{-3}$) | $\rho$ ($\Omega$cm) | $\mu$ (cm$^2$/Vs) | $T_C$ (K) |
|---|---|---|---|---|---|---|
| 5 | 15 | N | 8.19×10$^{16}$ | 1.06 | 72.0 | 120 |
| 10 | 15 | N | 8.39×10$^{16}$ | 1.28 | 58.2 | 80 |
| 20 | 15 | P | 6.54×10$^{17}$ | 0.35 | 27.3 | 40 |
| 30 | 10 | P | 6.15×10$^{17}$ | 0.14 | 72.6 | 110 |

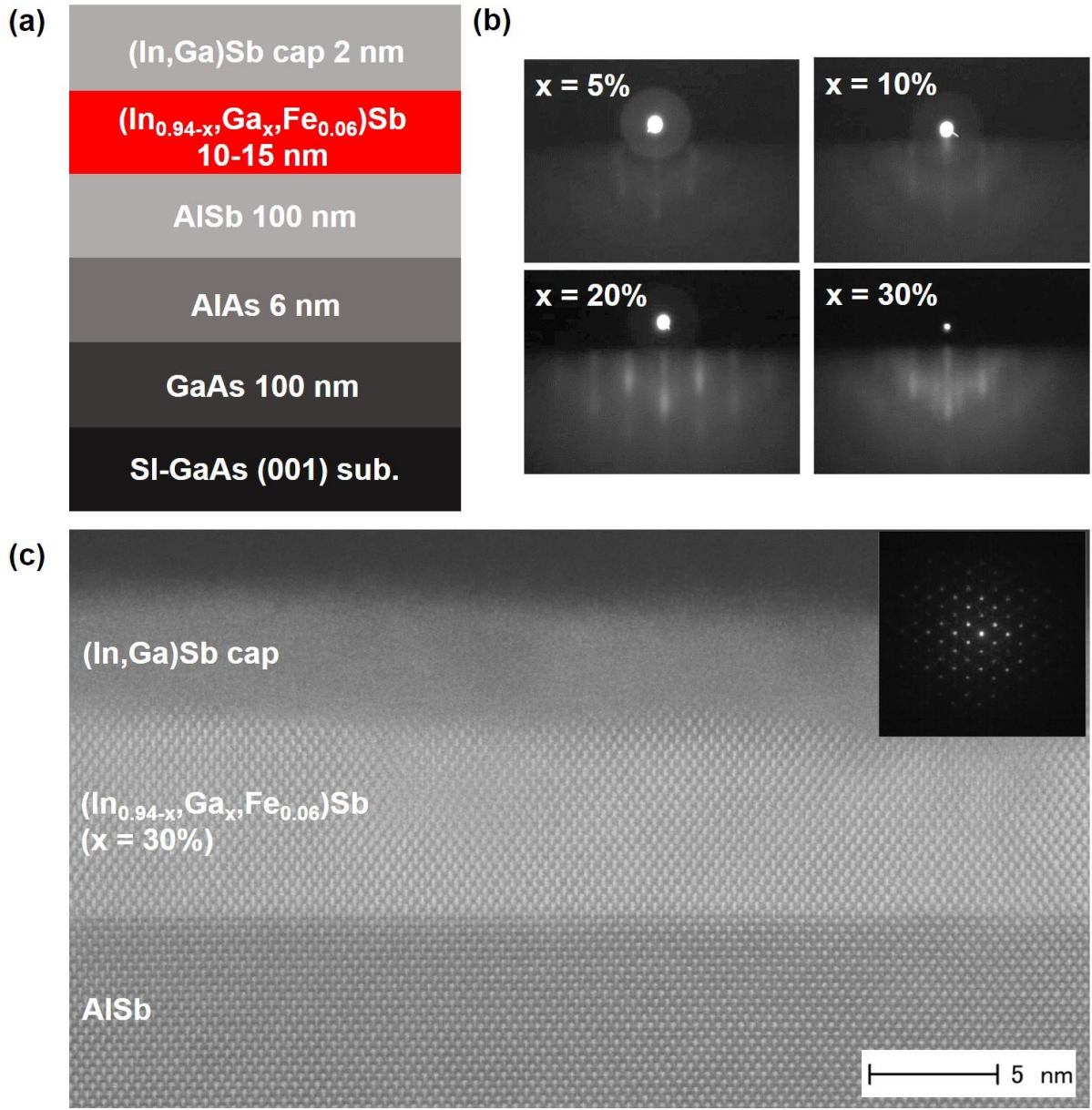

**Figure 1**. (**a**) Schematic structure of the samples studied in this work. (**b**) *In-situ* reflection high energy electron diffraction (RHEED) patterns along the [$\bar{1}$10] azimuth during the MBE growth of the (In$_{0.94-x}$,Ga$_x$,Fe$_{0.06}$)Sb ($x = 5 - 30\%$) layers. (**c**) Scanning transmission electron microscopy (STEM) lattice image of the (In$_{0.94-x}$,Ga$_x$,Fe$_{0.06}$)Sb ($x = 30\%$) sample, taken along the [110] direction. Inset shows the transmission electron diffraction (TED) pattern of the (In$_{0.94-x}$,Ga$_x$,Fe$_{0.06}$)Sb ($x = 30\%$) layer. These results indicate that the crystal structure is of zinc-blende type.

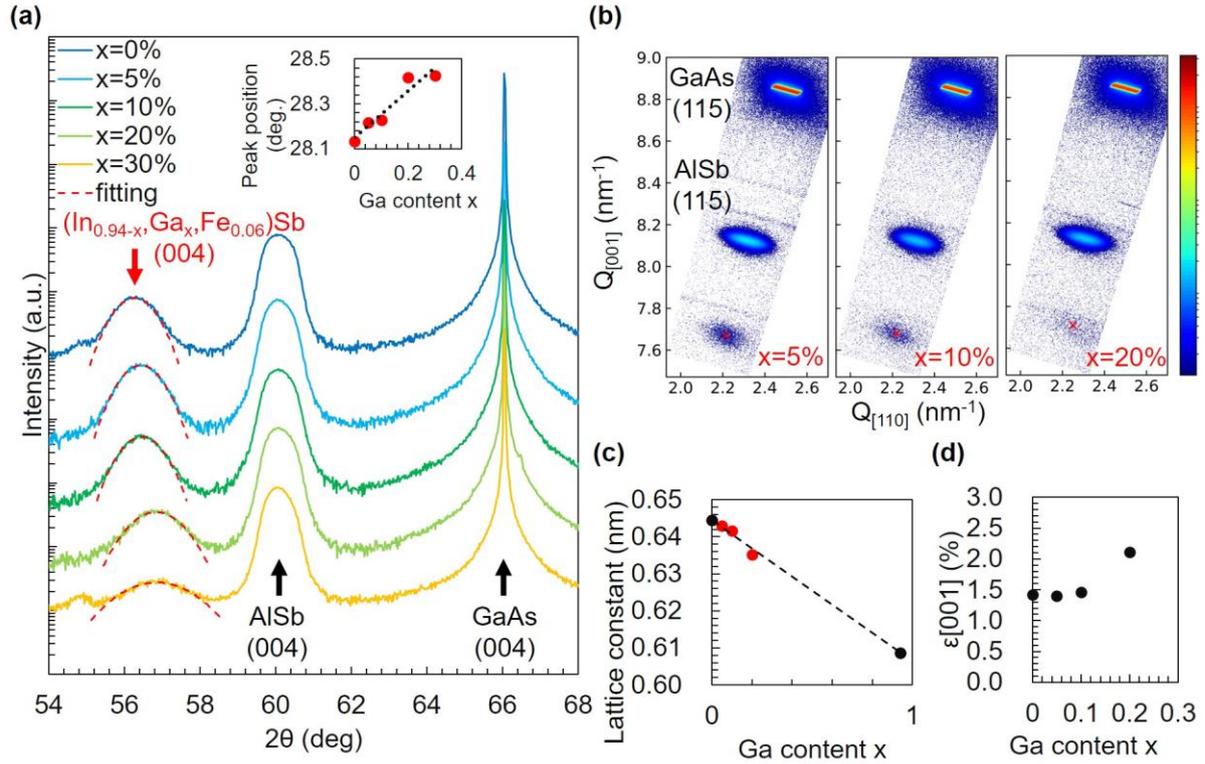

**Figure 2**. **(a)** X-ray diffraction (XRD) spectra of the samples studied in this work. The spectrum of an $(In_{0.94},Fe_{0.06})Sb$ ($x = 0\%$) sample is also shown as a reference. Gaussian fitting curves (red dot line) are used to estimate the (004) peak position of the $(In_{0.94-x},Ga_x,Fe_{0.06})Sb$ layer in each sample. Inset shows the estimated peak positions *vs*. Ga content $x$. **(b)** Reciprocal space mapping (RSM) of the $(In_{0.94-x},Ga_x,Fe_{0.06})Sb$ ($x = 5, 10, 20\%$) samples around the (115) reciprocal point. The red cross mark indicates the peak position of the $(In_{0.94-x},Ga_x,Fe_{0.06})Sb$ layer in each sample. **(c)** Ga content $x$ dependence of the intrinsic lattice constants $a$ of the $(In_{0.94-x},Ga_x,Fe_{0.06})Sb$ layers estimated from the peak positions in the RSM results, which follows the Vegard's law. **(d)** $x$-dependence of the out-of-plane strain $\epsilon_{[001]}$ in all the samples.

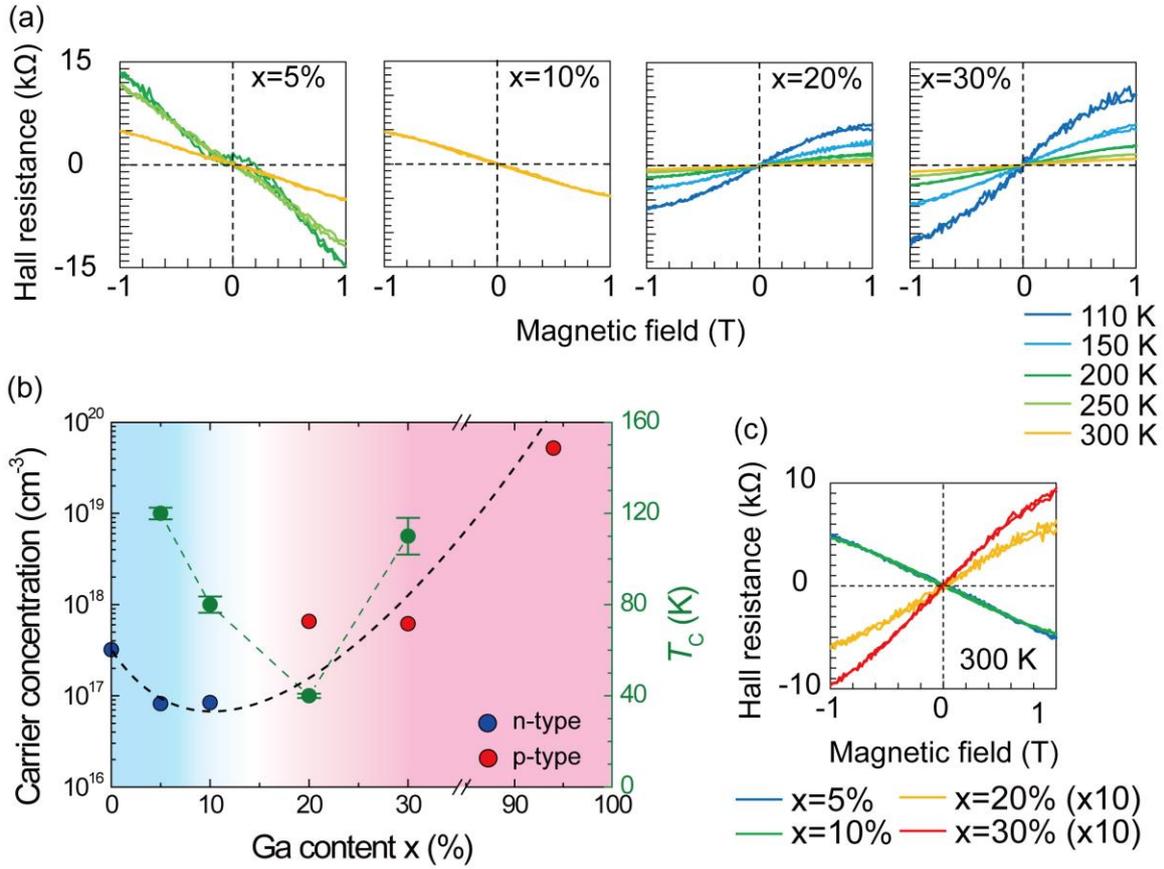

**Figure 3. (a)** Hall resistances of the $(In_{0.94-x},Ga_x,Fe_{0.06})Sb$ ($x = 5 – 30\%$) samples, measured at various temperatures. **(b)** Carrier concentration estimated at 300 K (blue and red circles, left axis) and $T_C$ (green circles, right axis) of the $(In_{0.94-x},Ga_x,Fe_{0.06})Sb$ layers as functions of the Ga content $x$. The carrier types and concentrations are estimated from the Hall coefficients at 300 K. The blue and red circles correspond to n-type and p-type conductions, respectively. A transition from n-type conduction (blue background region, $x < 10\%$) to p-type conduction (red background region, $x > 20\%$) is observed. Black dotted curve is plotted to show the trend of carrier concentration. **(c)** Hall resistances of the $(In_{0.94-x},Ga_x,Fe_{0.06})Sb$ samples measured at 300 K, in which the Hall resistance values of the samples with $x = 20$ and 30%, which are p-type, are magnified by 10 times for clear comparison with those in the n-type samples ($x = 5, 10\%$). The sign of the Hall coefficient at 300 K clearly switches from negative to positive when increasing $x$ over 10%.

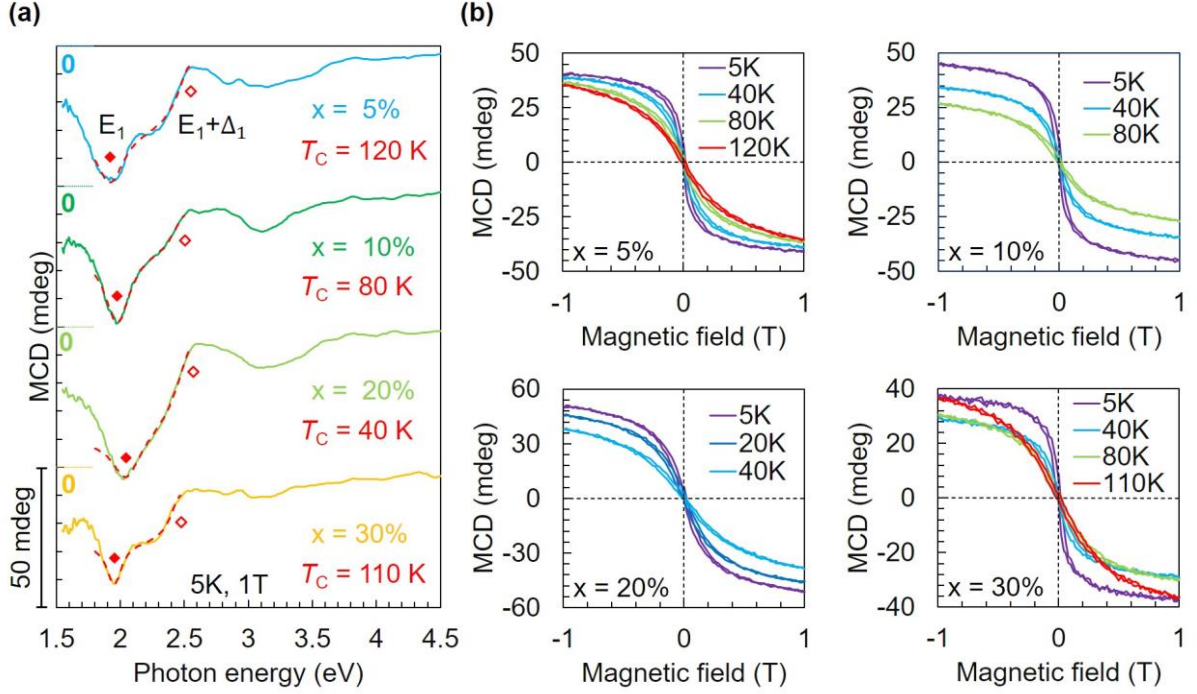

**Figure 4**. (a) Magnetic circular dichroism (MCD) spectra of the $(In_{0.94-x},Ga_x,Fe_{0.06})Sb$ ($x$ = 5 – 30%) layers at 5 K under a magnetic field $H$ of 1 T applied perpendicularly to the films, which are shifted vertically for clarity. We fit a sum of negative and positive Lorentzian fitting curves (red dot line) to the MCD peaks around 1.8 – 2.5 eV to estimate the critical point energies $E_1$ (red filled square) and $E_1+\Delta_1$ (red blank square) of each sample. (b) MCD – $H$ hysteresis curves of all the samples measured at a photon energy of 2.0 eV at various temperatures. We estimate $T_C$ in these samples from the Arrott plots of these MCD – $H$ curves.